\documentstyle[12pt]{article}
\newcommand{\referencestyle}{
\small
\abovedisplayskip=6pt
\belowdisplayskip=6pt
\vspace{12pt}}

\def\D{\Delta}

\def\t0{\tau_0}

\def\ben{\begin{eqnarray}}
\def\enn{\end{eqnarray}}
\def\ov{\over\displaystyle\phantom{|}}
\def\dst{\displaystyle\phantom{|}}
\def\l({\left(}
\def\r){\right)}
\def\ts{\tilde s_c}
\def\ave#1{\langle #1 \rangle}

\begin{document}
\include{epsf}
%\rightline{hep-ph/9705422}
%\rightline{CU-TP-819R/1997}
%\rightline{DRAFT}
%\bigskip
\begin{center}
{ \Large\bf
Higher Order Bose-Einstein Correlations for \\
Systems with Large Halo}
\end{center}
\medskip
\begin{center}
        T. Cs\"org\H o$^{1,2}$\footnote{E-mails:
                csorgo@sunserv.kfki.hu, csorgo@nt1.phys.columbia.edu }
\end{center}
%\medskip
\begin{center}
{\it
$^1$ Department of Physics, Columbia University, 10027 New York, N.Y.\\
$^2$ MTA KFKI RMKI, H--1525 Budapest 114, P.O. Box 49. Hungary \\
}
\end{center}
\medskip
\begin{abstract}
A generalization of the core/halo model to
the Bose-Einstein correlation function of any number of particles is
presented.  In particular, a simple prediction is obtained
for the intercept parameter of the $n$-particle Bose-Einstein
correlation functions for arbitrary large values of $n$.
\end{abstract}
\bigskip
\bigskip

{\it Introduction}.
Multi-particle correlations are studied extensively in high energy
collisions, with the aim of learning more about possible phase transitions
or self-similar fluctuations. Especially, the short-range
part of the two-particle correlation function,
is thought to carry information about the space-time structure based on a
quantum-statistical effect discovered  in refs.~\cite{gglp,hbt}.
One of the most  important driving force behind correlation studies
in high energy physics is the
possibility to measure space-time dimensions on the 10$^{-15}$ m and
10$^{-23}$ sec scale. See refs.~\cite{qm96,s96} for recent reviews
 on high energy heavy ion physics,
for summaries on correlation studies
in this field see refs.~\cite{gyu_ka,bengt,zajc,uli_rev}.

Multi-particle correlations are becoming experimentally well determinable in
present and planned heavy ion reactions at
at Brookhaven AGS, at CERN SPS,  at the Relativistic Heavy Ion Collider
(RHIC) and at CERN LHC both due to the very large number of the produced
particles and due to the dramatic increase of the data quality accessible
already for multidimensional correlation studies, ref.~\cite{janus}.

There has been a lot of development recently
in the theory of multi-particle Bose-Einstein correlations,
form partial coherence to multi-particle wave-packet models
and event generators with multi-particle symmetrization,
refs.~\cite{crak,ncramer,eggers,henning,plaser,zhang,lednicky_s96,jozsow}.
The purpose of the present Letter is to study the structure of
multi-particle Bose-Einstein correlations in one special case,
when the emission function can be well separated to a core and a halo part.
Such a scenario is  referred to as the
core/halo model~\cite{chalo}.
The formal study of the core/halo type of models first started with
numerical simulations,~\cite{halo1,pratt_csorgo,
lutp,RQMD,bowler,bow}.
The essential ideas necessary to formulate the
core/halo model were also published in ref.~\cite{halo2}, see refs.
{}~\cite{halo3,halo4} for further details.

In refs.~\cite{edge1,edge2,edge3} an Edgeworth expansion
method was proposed to characterize the non-Gaussian features
of the Bose-Einstein correlation functions as caused by interference
of decay product pions from long-lived resonances.
However, the data indicated surprisingly regular, simple
Gaussian correlation functions when the analysis was performed in
 two or three dimensions~\cite{NA44,NA35}.
This resulted~\cite{3d} in the re-formulation and simplification of the
theoretical aspects of the resonance-decay effects on the Bose-Einstein
correlation functions in a particularly transparent form,
with an emphasis on the interpretation and  possible utilization of the
momentum dependent intercept parameter of the two particle Bose-Einstein
correlation function in ref.~\cite{chalo}.
We present herewith the generalization of this core/halo model
to the $n$-particle case, utilizing the Wigner-function formalism.

In this formalism, the one-boson emission
is characterized by the emission function, $S(x,p)$.
Here $x = (t,{\bf r}\,) $ and $p = (E, {\bf p} \,) $
denote  four-vectors in space-time and in momentum space,
for particles with mass $m = \sqrt{ E^2 - {\bf p}^{\, 2} }$.
An auxiliary quantity is
\ben
\tilde S(\Delta k , K ) & = & \int d^4 x \,\,
                 S(x,K) \, \exp(i \Delta k \cdot x ),\label{e:ax}
\enn
where
$
\Delta k  = p_1 - p_2$, $K  = {(p_1 + p_2)/ 2}$
and $\Delta k \cdot x $ stands for the inner product
of the four-vectors. The invariant momenum distribution reads as
\ben
 E {\dst d n \ov d{\bf p} } & = & N_1({\bf p}) \,
= \, \tilde S(\Delta k = 0, K = p).\label{e:imd}
\enn
In this Letter we shall utilize the `hydrodynamical' normalization of the
Wigner-functions,
\ben
\int
{\dst d{\bf p}\ov E} d^4x S(x,p) & = &
\int {\dst d{\bf p}\ov E} N_1({\bf p})  \,\, = \,\, \ave{n},\label{e:nimd}
\enn
where $\ave{n}$ is the mean multiplicity.
The two-particle
 BECF-s are prescribed as
\ben
C_2({\bf p}_2,{\bf p}_2)
	& \simeq &
	1 +  {\displaystyle\strut
         \mid \tilde S(\Delta k , K) \mid^2 \ov
		\tilde S(0,{\bf p}_1) \tilde S(0,{\bf p}_2) },
	\,\, \simeq \,\,
	1 +  {\displaystyle\strut
         \mid \tilde S(\Delta k , K) \mid^2 \ov \mid \tilde S(0,K)\mid^2 },
\enn
 as was presented e.g. in ref. \cite{pratt_csorgo,zajc}.
The precision of the last approximation was estimated to be of the
order of 5 \% as discussed in Ref.~\cite{uli_l}.
In this Letter final state interactions are neglected
and a completely chaotic particle  emission is assumed.

One can show~\cite{rrr} that the higher order
Bose-Einstein correlation functions,
\ben
C_n({\bf p}_1,{\bf p}_2,...,{\bf p}_n) & = &
	{\dst N_n({\bf p}_1,{\bf p}_2,...,{\bf p}_n) \ov
	N_1({\bf p}_1) N_1({\bf p}_2) ... N_1({\bf p}_n) }
\enn
are given in terms of the Fourier-transformed Wigner-functions as
\ben
C_n({\bf p}_1,{\bf p}_2,...,{\bf p}_n) & = &
	{\dst \sum_{\sigma^n} \prod_{i=1}^n \tilde S(i,\sigma_i)
	\ov \prod_{i=1}^n \tilde S(i,i)} \,\, = \,\,
	\sum_{\sigma^n}  \prod_{i=1}^n {\dst  \tilde S(i,\sigma_i)
	\ov \tilde S(i,i)}  \,\, = \,\,
	\sum_{\sigma^n}  \prod_{i=1}^n  \tilde s(i,\sigma_i),
\enn
where the summation is over all the $\sigma^n$ permutations of $n$ indexes,
\ben
\tilde S(i,\sigma_i) \, = \, \tilde S(K_{i,\sigma_i},\Delta k_{i,\sigma_i}),
	& \mbox{\rm and} &
	\tilde s(i,\sigma_i) \, = \,
	{\dst \tilde S(i,\sigma_i)
	\ov  \tilde S(i,i) }, \\
	K_{i,\sigma_i} \, = \, {\dst p_i + p_{\sigma_i} \ov 2},
	& \mbox{\rm and} &
	\Delta k_{i,\sigma_i} \, = \, p_i -  p_{\sigma_i}.
\enn
Note the distinction between $\sigma^n$, which stands for the set of the
permutations of the numbers $1, 2, ..., n$ and the $\sigma_i$ (subscript $i$),
which indicates the number replacing the index $i$ in a given permutation from
the set $\sigma^n$. Note also that $\tilde s(i,\sigma_i)
= {\tilde S(\sigma_i,\sigma_i) \over \tilde S(i,i)}
\tilde s^*(\sigma_i,i) \ne \tilde s^*(\sigma_i,i) $,
 although the relationship
$\tilde S(i,\sigma_i) = \tilde S^*(\sigma_i,i)$ is satisfied.

In the core/halo model, the following assumptions are made :

\underline{\it Assumption 0:}
The emission function does not have a no-scale, power-law like structure.
This possibility was discussed and related to intermittency in
ref.~\cite{bialas}.

\underline{\it Assumption 1:} The bosons are emitted either from a {\it
central} part or
from the surrounding  {\it halo}. Their emission functions
are indicated by $S_c(x,p)$ and $S_h(x,p)$, respectively.
According to this assumption, the complete emission function can be written
 as
\ben
 S(x,p) = S_c(x,p) +  S_h(x,p),
\enn
using the hydrodynamic normalization of the Wigner functions.

\underline{\it Assumption 2:} We assume that the emission function which
characterizes
the halo changes on a scale $R_H$ which is larger
 than  $R_{max}\approx \hbar / Q_{min}$, the maximum length-scale
resolvable~\cite{chalo}
by the intensity interferometry microscope. However, the smaller central part
of size $R_c$ is
assumed to be resolvable,
 $ R_H > R_{max} > R_c $.
This inequality is assumed to be satisfied by all characteristic scales
in the halo and in the central part, e.g. in case the side, out or longitudinal
components~\cite{bertsch,lutp} of the correlation function are not identical.

\underline{\it Assumption 3:}  The momentum-dependent core fraction
$f_c(i) = N_c({\bf p}_i)/N_1({\bf p}_i)$ varies slowly on the relative
momentum scale given by the correlator of the core $\ts(1,2) \ts(2,1)$.

According to this smoothness assumption, which was only implicitly made
in ref.~\cite{chalo}, the relative momentum dependence
of the core fraction can be neglected on the relative momentum
scales on which the correlator of the core is changing.
If a typical core radius parameter in a given direction  is
indicated by $R_c$ then the slow variation of $f_c(i)$ implies
that $f_c({\bf p}_i) f_c({\bf p}_j) \simeq f_c^2({\bf K}_{i,j})$
if $|{\bf p}_i - {\bf p}_j| < \hbar/R_c$.

The emission function of the center and that of the halo are normalized
here as
\ben
\int d^4x {\dst d{\bf p}\ov E} S_c(x,p) \, = \, \ave{n}_c,
& \mbox{\rm and}&
\int d^4x {\dst d{\bf p}\ov E} S_h(x,p) \, = \, \ave{n}_h,
\enn
where the subscripts $c,h$ index the contributions by the central and the
halo parts, respectively.
In ref~\cite{chalo}, the above Assumptions were formulated with the
help of the Wigher-functions normalized to 1, and a core-fraction
$f_c = \ave{n}_c / \ave{n}$ was used, while in this paper
the {\it momentum-dependent} core fraction $f_c({\bf p})$ is utilized.
 One finds that
\ben
 N_1({\bf p}) \, =  \, N_c({\bf p}) + N_h({\bf p}),
& \mbox{\rm and}&
\ave{n} \, = \, \ave{n}_c + \ave{n}_h.
\enn
Note, that in principle the core as well as the halo part of
the emission function could be decomposed into more detailed contributions.
In case of pions and NA44 acceptance~\cite{csorgo-kiang},
one can separate the contribution of various long-lived resonances as
\ben
	S_{h}(x,p) \, = \, \sum_{r=\omega,\eta,\eta^{\prime},K^0_S}
		S^{(r)}_{halo}(x,p)
		& \mbox{\rm and} &
	N_h({\bf p}) \, = \,\sum_{r=\omega,\eta,\eta^{\prime},K^0_S}
		N^{(r)}_{halo}({\bf p}).
\enn
For our considerations, this separation is indifferent.
According to our assumptions, the Fourier - transformed Wigner-functions
characterizing the (full) halo part are narrower than the two-particle momentum
resolution $Q_{min}$, for which we typically may take the value of cca. 10 MeV.
It is important to keep in mind that the halo part of the emission function
is defined with respect to $Q_{min}$. For example, if $Q_{min} = 10-15$ MeV,
the decay products of the $\omega$ resonances can be taken as parts of
the halo~\cite{csorgo-kiang}, however,
should the experimental resolution {\it and} the error bars on the
measurable correlation funtion decrease {\it significantly} below
$\hbar/\Gamma_{\omega} = 8$ MeV, the decay products of the $\omega$
resonances would contribute to the core.
See refs.~\cite{csorgo-kiang,chalo} for a more complete discussion on
this specific topic.

The measured $\tilde S_h(i,\sigma_i)$ is thus vanishing if $i \ne \sigma_i$
at the given measured relative momentum $Q_{i,\sigma_i} > Q_{min}$.
Note that $Q_{i,\sigma_i}$ can be one, two-
or three-dimensional quantity, e.g. any reasonable measure of
$\Delta k_{i,\sigma_i}$. However, it is important to observe that
$\tilde S_h(i,i)$ contributes to the spectrum, and is not affected by
the {\it two-}particle momentum resolution:
\ben
 \tilde S_h(i,\sigma_i) & = & \delta_{i,\sigma_i} \tilde S(i,i),\\
\tilde s(i,\sigma_i) & = &
	\delta_{i,\sigma_i}  +
	(1 - \delta_{i,\sigma_i}) f_c(i) \tilde s_c(i,\sigma_i ) , \\
	f_c(i) & = & f_c({\bf p}_i) \, = \, N_c({\bf p}_i) / N_1({\bf p}_i).
\enn
Thus the Bose-Einstein correlation function
	$C_n(1,2, ... ,n)  =  C({\bf p}_1,{\bf p}_2, ... ,{\bf p}_n)$
	reads as
\ben
	C_n(1,2,...n) & = & \sum_{\sigma_n} \prod_{i=1}^n
	\left[ \delta_{i,\sigma_i} + ( 1 -  \delta_{i,\sigma_i}) f_c(i)
	\tilde s_c(i,\sigma_i)
	\right]
\enn
Let us introduce $\rho^n$ to stand for those permutations of $(1,...,n)$
which are mixing  {\it all} the  numbers from 1 to $n$ and let us indicate by
$\rho_i$ the value which is  replaced by $i$ in a given permutation belonging
to the set of permutations $\rho^n$.
(Superscript indexes a set of
permutations, subscript stands for a given value).
Then we have $\rho_i \ne i$ for all values of $i = 1,..., n$,
while
$\sigma_i = i$ type of replacements are allowed.

Thus the general expression for $C_n(1,...,n)$ reads as
\ben
C_n(1,...,n) & = & 1 + \sum_{j = 2}^n
	\sum_{i_1, ..., i_j = 1}^{\null \,\,\, n \,\,\, _{\prime}}
	\sum_{\rho^n} \prod_{k=1}^j
	f_c(i_k) \tilde s_c(i_k,i_{\rho_k}).
	\label{e:fmix}
\enn
Here $\sum'$ indicates that the summation should be taken over  those set of
values of the indices which do not contain any value more than once.
%corresponding to the nature of the fully mixing permutations included into
%$\rho^n$.

Let us explicitly write out the above expression until the $l = 4$ terms:
\ben
C_n(1,...,n)  & = &
	1 + \sum_{i\ne j = 1}^n f_c(i) f_c(j)
		\left[\tilde s_c(i,j) \tilde s_c(j,i) \right]
	+
	\nonumber \\
\null & \null & \, + \,
	\sum_{i\ne j \ne k= 1}^n \! f_c(i) f_c(j) f_c(k)
	\left[ \tilde s_c(i,j) \tilde s_c(j,k) \tilde s_c(k,i) +
	 \tilde s_c(i,k) \tilde s_c(k,j) \tilde s_c(j,i)\right]
	+
	\nonumber \\
\null & \null & \, + \,
	\sum_{i\ne j \ne k \ne l = 1}^n \!\! f_c(i) f_c(j) f_c(k) f_c(l) \times
	\nonumber \\
\null & \null & \, \times \,
	\left[
	\tilde s_c(i,j) \tilde s_c(j,k) \tilde s_c(k,l)	\tilde s_c(l,i)	+
	\tilde s_c(i,j) \tilde s_c(j,l) \tilde s_c(l,k)	\tilde s_c(k,i)	+
	\right.
	\nonumber \\
\null & \null & \, + \,
	\tilde s_c(i,k) \tilde s_c(k,j) \tilde s_c(j,l)	\tilde s_c(l,i)	+
	\tilde s_c(i,k) \tilde s_c(k,l) \tilde s_c(l,j)	\tilde s_c(j,i)	+
	\nonumber \\
\null & \null & \, + \,
	\tilde s_c(i,l) \tilde s_c(l,j) \tilde s_c(j,k)	\tilde s_c(k,i)	+
	\tilde s_c(i,l) \tilde s_c(l,k) \tilde s_c(k,j)	\tilde s_c(j,i)	+
	\nonumber \\
\null & \null & \, + \,
	\tilde s_c(i,j) \tilde s_c(j,i) \tilde s_c(k,l)	\tilde s_c(l,k)	+
	\tilde s_c(i,k) \tilde s_c(k,i) \tilde s_c(j,l)	\tilde s_c(l,j)	+
	\nonumber \\
\null & \null & \, + \,
	\left.
	\tilde s_c(i,l) \tilde s_c(l,i) \tilde s_c(j,k)	\tilde s_c(k,j)
	\right]
	+ ...... \label{e:chn}
\enn
 There are 44  (and 265) terms in the next part which fully mixes
 5 (and 6) different indexes for any given fixed value
of $i \ne j \ne k \ne l \ne m (\ne n)$, respectively.

For the two-particle case, we recover the earlier result given in
Ref.~\cite{chalo}. Let us proceed very carefully at this point,
so that the role of Assumption 3 could be made transparent.
If only two particles are present, the above equation reduces to
\ben
	C_2(1,2)  & = & 1 + f_c(1) f_c(2) \ts(1,2) \ts(2,1).
\enn
If Assumption 3 is also satisfied by some experimental data set,
then the resulting formula becomes particularly simple:
\ben
{ C(\D k_{12}, K_{12}) }
	 & = & 1 +
      \lambda_*(K_{12}) {\dst \mid \tilde S_c( \D k, K) \mid^2 \ov
                             \tilde S_c( 0, p_1) \tilde S_c(0,p_2)},
	 \,  \simeq \, 1 +
      \lambda_*(K_{12}) {\dst \mid \tilde S_c( \D k, K) \mid^2 \ov
                            | \tilde S_c( 0, K_{12})|^2}, \\
                           \label{e:lamq}
\enn
where the effective intercept parameter $\lambda_*(K_{12})$
is given as
$
	\lambda_*(K_{12})  =
		\left[N_c(K_{12}) / N_1(K_{12}) \right]^2.
$
As emphasized in Ref.~\cite{chalo}, this {\it effective} intercept parameter
( $\ne$ exact intercept parameter at $Q = 0$ MeV)
shall in general depend on the
mean momentum of the observed boson pair,
which within the errors of $Q_{min}$ coincides
with any of the on-shell four-momentum $p_1$ or $p_2$.
Thus one obtains the core/halo interpretation of the two-particle
correlation function: the intercept parameter $\lambda_*(K)$ measures
the momentum dependent core fraction and the relative momentum dependent
part of the two-particle correlation function is determined by the
core. Note, that in high energy heavy ion collisions the momentum
dependence of the $\lambda_*(K)$ parameter is very weak, actually,
within the errors $\lambda_*(K)$ is constant for the NA44 data
analyzed in ref.~\cite{chalo}. However, the validity of this
{\it Assumption 3} has to be checked experimentally for each
data set, by determining the momentum dependence of the
$\lambda(K)$ parameter of the two-particle correlation function.

Within the core/halo picture, the
 $n$-particle correlation function  has a simple form if
all the $n$ momenta are approximately equal, i.e.
$\mid \Delta k_{i,\sigma_i} \mid \le Q_{min}$ for all
$i \ne \sigma_i$, and this situation is denoted by
${\bf p}_1 \simeq {\bf p}_2 \simeq ... \simeq {\bf p}_n \simeq {\bf P}$.
One obtains that
$$
C_n({\bf p}_1 \simeq {\bf p}_2 \simeq ... \simeq {\bf p}_n \simeq {\bf P})
	 =  1 + \sum_{j = 1}^n f_c({\bf P})^j
         \left( \null^{\dst n}_{\dst  j}  \right) \alpha_j
$$
where $\alpha_j$ stands for the number of fully mixing permutations of $j$
indexes, i.e. the number of permutations in $\rho^j$.
The first few values of $\alpha_j$ are given as
\ben
	\alpha_1 \, = \, 0, \qquad
	\alpha_2 \, = \, 1, \qquad
	\alpha_3 \, = \, 2, &\quad &
	\alpha_4 \, = \, 9, \qquad
	\alpha_5 \, = \, 44,\qquad
	\alpha_6 \, = \, 265.
\enn
These values can be obtained from a recurrence relation, as follows.
Let us indicate the number of permutations that completely mix  exactly
$j$ non-identical elements by $\alpha_j$. There are
exactly $\left( \null^{\dst n}_{\dst  j} \right)$ different
ways to choose $j$ different elements from among $n$ different elements.
Since all the $n!$ permutations can be written as a sum over the
fully mixing permutations, the counting rule yields
$
	n!  =  1 + \sum_{j = 1}^n
	\left(  \null^{\dst n}_{\dst  j} \right) \alpha_j,
$
which can be rewritten as a recurrence relation for $\alpha_j$:
\ben
	\alpha_n & = & n! - 1 - \sum_{j = 1}^{n - 1}
	{\left( \null^{\dst n}_{\dst  j}  \right)} \alpha_j.
	\label{e:alp}
\enn
We have the following explicit expressions for the first few intercept
parameters:
\ben
	\lambda_{*,2}({\bf P}) & = & f_c({\bf P})^2
		\label{e:l2} \\
	\lambda_{*,3}({\bf P}) & = & 3 f_c({\bf P})^2 + 2 f_c({\bf P})^3 \\
	\lambda_{*,4}({\bf P}) & = & 6 f_c({\bf P})^2 + 8 f_c({\bf P})^3
					+ 9 f_c({\bf P})^4\\
	\lambda_{*,5}({\bf P}) & = & 10 f_c({\bf P})^2 + 20 f_c({\bf P})^3
		+ 45 f_c({\bf P})^4 + 44 f_c({\bf P})^5 \\
	\lambda_{*,6}({\bf P}) & = & 15 f_c({\bf P})^2 + 40 f_c({\bf P})^3
		+ 135 f_c({\bf P})^4 + 264 f_c({\bf P})^5  +
		265 f_c({\bf P})^6 \label{e:l6}
\enn
In general, the intercept parameter of the $n$-particle
correlation function reads as
\ben
	\lambda_{*,n}({\bf P}) & = & \sum_{j = 1}^n f_c({\bf P})^j
         \left( \null^{\dst n}_{\dst  j}  \right) \alpha_j
\enn
where $\alpha_j$ is defined by the recurrence given in Eq.~(\ref{e:alp}).
These expressions then relate the effective intercept parameter of the
$n$-particle correlation function to the effective intercept parameter
of the two-particle correlation function, and could be thus checked
experimentally. For that type of test, a few warnings should be made:
The intercept parameters $\lambda_{*,n}({\bf P})$ explicitly depend on the
momentum of the particles in the region where all the $n$ particles have
approximately the same momentum. Thus averaging over the transverse mass and
the rapidity of the particles to improve statistics is in principle not
allowed. Further, the effective intercept parameter $\lambda_{*,n}({\bf P})$
{\it differs} from the
{\it exact} analytical value of the $n$-particle correlation
function which is $n!$ in our picture, due to the fact that
we have neglected possible partial coherence of the particle emitting source.
Our motivation for ignoring possible partial
coherence was to find a relatively simple limiting case, and also
we are aware of the fact that the pure quantum statistical relationship
between the second order and the higher order correlation functions
has been shown not to be consistent with the available UA1 data~\cite{eggers}.
In case of heavy ion reactions, resonance halo seems to be describing
the drop of the effective intercept parameter from the value of 2
both in numerical simulations~\cite{RQMD} and in analytical approaches
{}~\cite{chalo,henning}. Finally, we stress that the values for the effective
intercept parameters $\lambda_{*,n}({\bf P})$
 may depend on the two-particle momentum resolution
$Q_{min}$, on the error distribution on the correlation function,
 as well as on the Gaussian or non-Gaussian structure
for the Fourier-transformed Wigner-function of the core,
 $\tilde S_c(K,\Delta k)$.

Note also that the general result for the correlation function in
the core/halo model coincides with  a particular limiting
case of the expressions obtained by
Cramer and Kadija for the correlation functions
of the order 1, 2, 3, 4 and 5, namely, the case when particle
mis-identification (contamination) is taken into account but
the source is assumed to have no partially coherent component.
 Thus our results can be applied also to the evaluation of
the intercept parameter of higher order
correlation functions for data contaminated
by unidentified or errorneously identified particles.

{\it Prediction of higher order intercept parameters for
NA44 data}.
Equations (\ref{e:l2}-\ref{e:l6}) can also be used to predict the
intercept parameter of higher order correlation functions
from the measured intercept parameter of the
second order correlation function. The published NA44 correlation
data indicate that $\lambda_{2,*} \simeq 0.52 \pm 0.02$ approximately
independently of the transverse mass for both the low $p_t$
and the high $p_t$ data sample for pion pairs
in a central $S + Pb$ reaction at 200 A GeV.
If we interpret this $\lambda_{2,*} $ in the core/halo model, we obtain
that $f_c = 0.72$ is approximately independent of the transverse mass
in the NA44 acceptance range. Thus, the core/halo model
predicts the following measurable intercept parameters for the
higher order correlation functions of identified pions in
200 AGeV $S + Pb$ reactions in the NA44 acceptance:
\ben
	\lambda_{*,3}({\bf P}) \, = \, 2.3 \qquad
	\lambda_{*,4}({\bf P}) \, = \, 8.6  & \qquad  &
	\lambda_{*,5}({\bf P}) \, = \, 33.4 \qquad
	\lambda_{*,6}({\bf P}) \, = \, 148.0
\enn
It should be emphasized that the NA44 data constitute a
good sample to test the core/halo model on higher order correlation functions,
given the experimentally observed momentum independence of the $\lambda_{2,*}$
parameter~\cite{NA44}.

{\it In summary,} we have evaluated the higher order correlation functions
in the core/halo model analytically. A closed form of the correlation function
of arbitrary high order is given by eq.~(\ref{e:fmix})
in terms of the momentum-dependent
core fractions and in terms of permutations that completely mix a set of
indexes such that $\rho_i \ne i $. A recurrence relation has been found
which allows for the evaluation of the measured intercept parameter of the
$n$-particle correlation function for arbitrary high orders in a very
efficient manner. We emphasized that the separation of the core from the
halo is dependent on the relative momentum resolution of the experiment
and pointed out a formal analogy between the correlation functions
of the  core/halo model and those of completely chaotic
sources with particle mis-identification.  Our results allow for a prediction
of the higher order correlation functions if the basic building block,
the amplitude $\tilde s(i,j)$
is determined experimentally as a
function of the mean and the relative momentum of the particle pair.
If this quantity is real, it can be determined from a detailed analysis
of the two-particle correlation function. If the imaginary part of $\tilde
s(i,j)$ is not negligible, a simultaneous analysis of second and third order
correlation functions is necessary to extract the building
block of higher order correlations.
%\vfill\eject

{\it Acknowledgments:}
Cs.T. would like to thank B. L\"orstad and J. Schmidt-S{\o}rensen
for discussions and collaboration in the  early phase of this work.
Thanks are due to D. Miskowiec for stimulating
discussions,  Gy\"orgyi and M. Gyulassy as well as
Bengt L\"orstad for their kind hospitality at Columbia University
and at University of Lund.  This work was supported by the Hungarian
NSF  under Grants  No. OTKA F4019, T024094, T016206
by the USA-Hungarian Joint  Fund by grant  MAKA 378/93
and by an Advanced Research Award from the Fulbright Foundation.

\null
\vfill
\eject
\end{document}